\documentclass[twocolumn]{aastex63}
\usepackage{natbib}
\usepackage{amsmath}
\usepackage{bm}

\shortauthors{Saha, Chakrabarty \& Sengupta}
\shorttitle{Multi-band transit follow up of five hot-Jupiters}

\begin{document}

\title{MULTI-BAND TRANSIT FOLLOW UP OBSERVATIONS OF FIVE HOT-JUPITERS WITH CRITICAL NOISE TREATMENTS: IMPROVED PHYSICAL PROPERTIES}

\author[0000-0001-8018-0264]{Suman Saha}
\affiliation{Indian Institute of Astrophysics, II Block, Koramangala, Bengaluru, India}
\affiliation{Pondicherry University, R.V. Nagar, Kalapet, Puducherry, India}

\correspondingauthor{Suman Saha}
\email{suman.saha@iiap.res.in}

\author[0000-0001-6703-0798]{Aritra Chakrabarty}
\affiliation{Indian Institute of Astrophysics, II Block, Koramangala, Bengaluru, India}
\affiliation{University of Calcutta, Salt Lake City, JD-2, Kolkata, India}

\author[0000-0002-6176-3816]{Sujan Sengupta}
\affiliation{Indian Institute of Astrophysics, II Block, Koramangala, Bengaluru, India}

\accepted{for publication in The Astronomical Journal}

\begin{abstract}
	
The most challenging limitation in transit photometry arises from the noises in the photometric signal. In particular, the ground-based telescopes are heavily affected by the noise due to perturbation in the Earth's atmosphere. Use of telescopes with large apertures can improve the photometric signal-to-noise ratio (S/N) to a great extent. However, detecting a transit signal out of a noisy light curve of the host star and precisely estimating the transit parameters call for various noise reduction techniques. Here, we present multi-band transit photometric follow-up observations of five hot-Jupiters e.g., HAT-P-30 b, HAT-P-54 b, WASP-43 b, TrES-3 b and XO-2 N b, using the 2m Himalayan Chandra Telescope (HCT) at the Indian Astronomical Observatory, Hanle and the 1.3m J. C.  Bhattacharya Telescope (JCBT) at the Vainu Bappu Observatory, Kavalur. Our critical noise treatment approach includes techniques such as Wavelet Denoising and  Gaussian Process regression, which effectively reduce both time-correlated and time-uncorrelated noise components from our transit light curves. In addition to these techniques, use of our state-of-the-art model algorithms have allowed us to estimate the physical properties of the target exoplanets with a better accuracy and precision compared to the previous studies.

\end{abstract}


\section{Introduction}

The transit photometry serves as one of the most important methods in the context of exoplanet detection and characterisation. This method helps us to determine several physical parameters of a transiting exoplanet, e.g., the radius, the inclination angle of the planetary orbit with respect to our line of sight (LOS), and the semi-major axis. However, a prior knowledge of the stellar radius is necessary for the estimation of these parameters. Modelling the transit light curves also allows us to determine the limb-darkening properties of the parent stars. Moreover, the orbital inclination angle estimated by using the transit method can be combined with the radial-velocity measurements, if available, and a prior knowledge of stellar mass to precisely estimate the mass of those exoplanets \citep{2019AJ....158...39C, southworth07}.

\begin{deluxetable*}{LCCCCC}
	\tablecaption{Adopted physical properties of the Exoplanets and their host-stars\label{tab:tab1}}
	\tablewidth{0pt}
	\tablehead{ & \colhead{HAT-P-30 b} & \colhead{HAT-P-54 b} & \colhead{WASP-43 b} & \colhead{TrES-3 b} & \colhead{XO-2 N b}}
	\startdata
	P\;[days] & 2.810595\pm0.000005 & 3.799847\pm0.000014  &  0.813475\pm0.000001 & 1.30618608\pm0.00000038 & 2.61585922\pm0.00000028 \\
	K_{RV}\;[m\:s^{-1}] & 88.1\pm3.3 & 132.6\pm4.9 & 551.0\pm3.2 & 378.4\pm9.9 & 90.17\pm0.82 \\
	R_\star\;[R_\sun] & 1.215\pm0.051 & 0.617\pm0.013 & 0.6506\pm0.0054 & 0.829\pm0.022 & 0.998\pm0.033 \\
	M_\star\;[M_\sun] & 1.242\pm0.041 & 0.645\pm0.020 & 0.688\pm0.037 & 0.928\pm0.048 &  0.96\pm0.05 \\
	T_{eff\star}\;[K] & 6304\pm88 & 4390\pm50 & 4500\pm100 & 5650\pm75 & 5332\pm57 \\
	\text{Sources} & \text{\citet{2011ApJ...735...24J}} & \text{\citet{2015AJ....149..149B}} & \text{\citet{2011AA...535L...7H}}, & \text{\citet{2009ApJ...691.1145S}}, & \text{\citet{2015AA...575A.111D}} \\
	&&& \text{\citet{2017AA...601A..53E}} & \text{\citet{2011ApJ...726...94C}}
	\enddata
\end{deluxetable*}

The photometric observations obtained from the ground-based telescopes are heavily affected by the turbulence in the Earth's atmosphere. This significantly adds to the overall noise in the observed signal. Moreover, if the ground-based survey telescopes used for the detection of new transiting exoplanets is small, it gives rise to a poor signal-to-noise ratio (S/N) in the observed transit signals. Therefore, repeated follow-up observations  are very important in order to estimate the physical properties of the confirmed exoplanets with a good accuracy and precision. Repeated follow-up observations with telescopes of larger aperture can result in high S/N in the transit light curves causing small error-bars in the light-curves. Further, in order to achieve an improved accuracy and precision in the values of the estimated transit parameters, application of critical noise reduction techniques is essential to reduce the fluctuations prevailing in the transit light curves. The results from transit follow-ups spanning over a large period of time can be used for the studies of planetary dynamics and may reveal the presence of any undiscovered planetary mass objects in those systems \citep{nesvorny2012detection, 2015ApJ...810L..23J, gillon17, patra17, maciejewski18}.

Our ongoing project involves the photometric follow-up of transiting exoplanets using multiple photometric bands and analysis using noise reduction techniques for an improved estimation of the physical properties. For photometric follow-up, we use the 2m Himalayan Chandra Telescope (HCT) at the Indian Astronomical Observatory, Hanle and the 1.3m J. C. Bhattacharya Telescope (JCBT) at the Vainu Bappu Observatory, Kavalur. \cite{2019AJ....158...39C} (hereafter, CS19) have reported the results from their observations of five hot Jupiters, namely WASP-33 b, WASP-50 b, WASP-12 b, HATS-18 b and HAT-P-36 b using the same telescopes as a part of this project and have addressed to the different sources of noise to improve the accuracy and precision of the estimated parameters. CS19 have segregated the noises according to their sources and spatial scales of span, and employed different techniques for the treatment of different kinds of such noises, including Wavelet Denoising and Gaussian Process regression.

As a continuation of this project, we have followed up five more hot Jupiters using the same telescopes but in multi-wavelength bands. These are  HAT-P-30 b \citep[e.g.,][]{2011ApJ...735...24J}, HAT-P-54 b \citep[e.g.,][]{2015AJ....149..149B}, WASP-43 b \citep[e.g.,][]{2011AA...535L...7H, 2017AA...601A..53E}, TrES-3 b \citep[e.g.,][]{2009ApJ...691.1145S, 2011ApJ...726...94C} and XO-2 N b \citep[e.g.,][]{2015AA...575A.111D}. The multi-band observations of transits enable us to estimate the wavelength dependent physical parameters, such as planetary radius, with a better accuracy corresponding to each photometric band.

One of the most prominent noise component in the photometric light-curves is the time-uncorrelated noise (white noise), which consists of both photon noise and the fluctuations in the light curve due to the small spatial scale variability in the transparency of Earth's atmosphere (CS19), such as atmospheric scintillation \citep{2015MNRAS.452.1707O, 2019MNRAS.489.5098F}. Most of the pre-processing techniques (such as binning and Gaussian smoothing) that can reduce the effect of these time-uncorrelated noise components also tend to distort the shape of the transit signal. CS19 have demonstrated that the Wavelet Denoising technique \citep{Donoho1994IdealDI, 806084, WaveletDenoise2012, 2018AA...619A..86D, cubillos17, waldmann14} can be used to reduce the time-uncorrelated fluctuations in the light-curves without distorting the transit signal and improves the precision of the estimated physical parameters (see Table 3 and Table 4 of CS19) to a great extent. Wavelet denoising technique also reduces the outliers in the light-curves due to cosmic ray events. In the present work, we have used the same technique to pre-process the transit light curves. However, while applying this technique in the present studies, we have made some appropriate upgradation. This was essential in order to handle multi-band data sets and a greater number of free parameters.

\begin{deluxetable*}{lccccCC}
	\tablecaption{Details of the photometric observations\label{tab:tab2}}
	\tablewidth{0pt}
	\tablehead{\colhead{Target Name} & \colhead{Date} & \colhead{Telescope} & \colhead{Instrument} & \colhead{Filter} & \colhead{No. of frames} & \colhead{S/N (Mean)}}
	\startdata
	HAT-P-30 b & 2020-02-05 & JCBT & UKATC & V & 42 & 747.92\\
	& 2020-02-05 & HCT & HFOSC & I & 315 & 2002.55\\
	& 2020-02-22 & HCT & HFOSC & R & 233 & 1508.76\\
	& 2020-03-10 & JCBT & UKATC & I & 50 & 482.48\\
	& 2020-03-24 & JCBT & UKATC & V & 22 & 747.37\\
	HAT-P-54 b & 2020-01-15 & JCBT & UKATC & I & 24 & 570.82\\
	& 2020-02-03 & JCBT & UKATC & I & 56 & 410.26\\
	& 2020-02-03 & HCT & HFOSC & V & 160 & 1001.14\\
	& 2020-02-22 & HCT & HFOSC & R & 93 & 1544.17\\
	WASP-43 b & 2019-03-06 & JCBT & UKATC & V & 85 & 379.06\\
	& 2019-04-02 & JCBT & UKATC & V & 93 & 390.22\\
	& 2019-04-11 & JCBT & UKATC & R & 54 & 606.31\\
	& 2019-12-31 & HCT & HFOSC & I & 261 & 1264.00\\
	& 2020-02-05 & HCT & TIRSPEC & J & 232 & 226.06\\
	& 2020-02-14 & JCBT & UKATC & R & 31 & 662.59\\
	TrES-3 b & 2019-02-18 & JCBT & UKATC & I & 96 & 283.19\\
	& 2019-05-03 & JCBT & UKATC & R & 50 & 337.95\\
	& 2020-03-31 & JCBT & UKATC & I & 26 & 537.23\\
	& 2020-09-21 & HCT & HFOSC & V & 203 & 1447.40\\
	XO-2 N b & 2020-01-28 & JCBT & UKATC & I & 54 & 710.21\\
	& 2020-02-05 & HCT & HFOSC & V & 292 & 1789.22\\
	& 2020-02-18 & JCBT & ProEM & I & 40 & 782.06\\
	& 2020-02-18 & HCT & HFOSC & R & 622 & 1261.56\\
	& 2020-03-23 & JCBT & UKATC & I & 27 & 699.05
	\enddata
\end{deluxetable*}

Another important noise component in the photometric signals is the time-correlated noise (red noise). We have reduced the large temporal scale red noise due to various instrumental and astrophysical effects by using the baseline correction method. On the other hand, the time-correlated fluctuations in the light curves of short temporal scale are due to the small spatial scale variations that affect each object on a frame differently (CS19). The major sources of this kind of red noise is the small-scale activity and pulsation of the host stars. Following CS19, we address this red noise component by using Gaussian Process (GP) regression method  \citep{2006gpml.book.....R, 2015ApJ...810L..23J, 2019MNRAS.489.5764P, 2020AA...634A..75B} to model it simultaneously while modelling for the transit signal.

In order to model the transit light curves, we have used the analytical formalism provided by \citet{2002ApJ...580L.171M}, which also incorporates the limb-darkening effect of the host-star using the quadratic limb darkening law. Following CS19, we have used the Markov Chain Monte Carlo (MCMC) method with the Metropolis-Hastings algorithm \citep{1970Bimka..57...97H} while modelling the transit light curves simultaneously along with the GP regression of the red noise. Although, the MCMC sampling technique is computationally expensive, it is extremely effective for modelling any noisy signal with a large number of free parameters.

Unlike the work by CS19, in the present work, we have set the radius of a planet as a free parameter for each wavelength band corresponding to each filter used. This allows us to get a coarse estimation of the band-dependent radius of the planets. This can be helpful in characterising the planets by studying the broad atmospheric features, if any, in those ultra low-resolution transit spectra detected with the help of the model transmission spectra in optical and in near infrared regions \citep{sengupta20, chakrabarty20}.

For the simultaneous handling of multiple observed photometric data from different nights of observation and a smooth and streamlined implementation of all the state-of-the-art techniques for reduction, photometry, processing, and modelling of the light curves, we have used  the software package developed by us. This Python-based package uses a semi-automated approach and is functionally not much different from the pipeline developed and used by CS19. All the steps in this package aim at a precise estimation of the planetary properties from the transit light curves, which are at the same time robust, accurate, and reliable.

In section \ref{sec:obs}, we have described our observational details. In section \ref{sec:analysis}, we have detailed our analysis and modelling techniques. In section \ref{sec:dis}, we have discussed the significance of our results and in section \ref{sec:con}, we have concluded our study.

\begin{figure*}[t]
	\centering
	\includegraphics[width=0.8\linewidth]{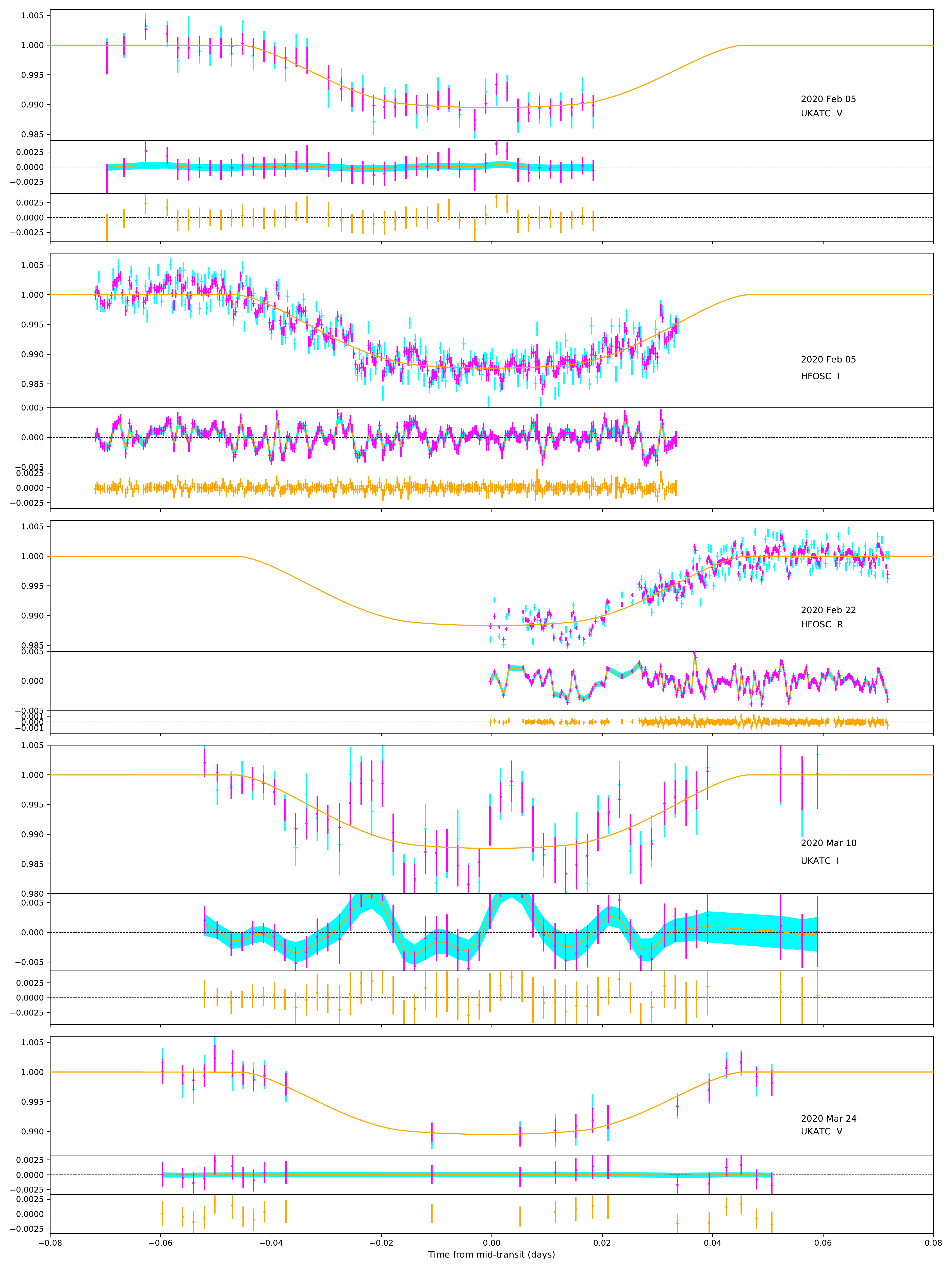}
	\caption{Observational and modelled light curves for HAT-P-30 b. For each observed transit event (see Table \ref{tab:tab2}), the observation date, the instrument, and the photometric filter used are mentioned. Top: the unprocessed light curve (cyan), light curve after Wavelet Denoising (magenta), the best-fit transit model (orange). Middle: the residual after modelling without GP regression (magenta), the mean (orange) and 1-$\sigma$ interval (cyan) of the best-fit GP regression model. Bottom: mean residual flux (orange).}
	\label{fig:fig1}
\end{figure*}

\begin{figure*}[t]
	\centering
	\includegraphics[width=0.8\linewidth]{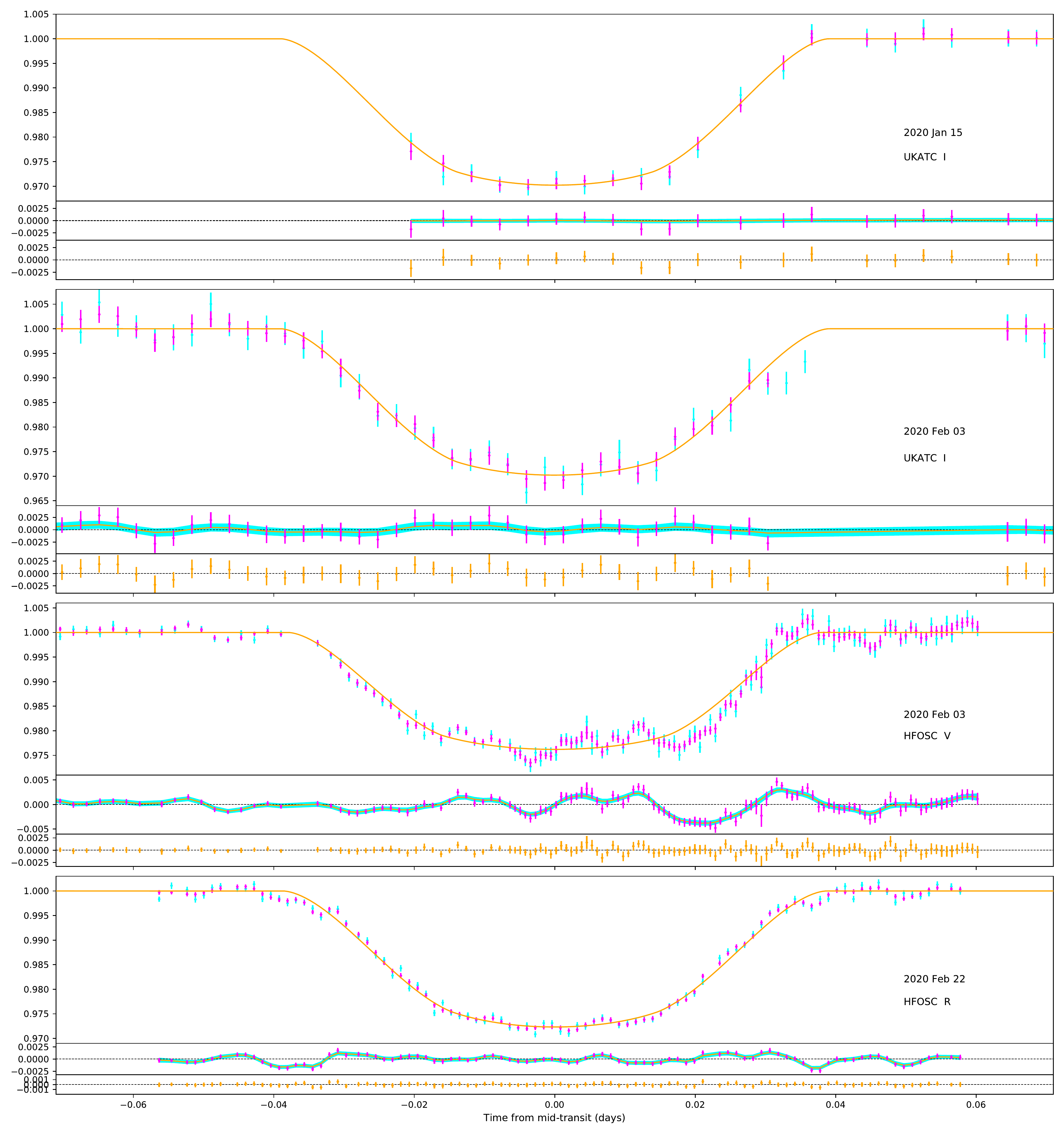}
	\caption{Same as Figure \ref{fig:fig1}, but for HAT-P-54 b}
	\label{fig:fig2}
\end{figure*}

\begin{figure*}[t]
	\centering
	\includegraphics[width=0.8\linewidth]{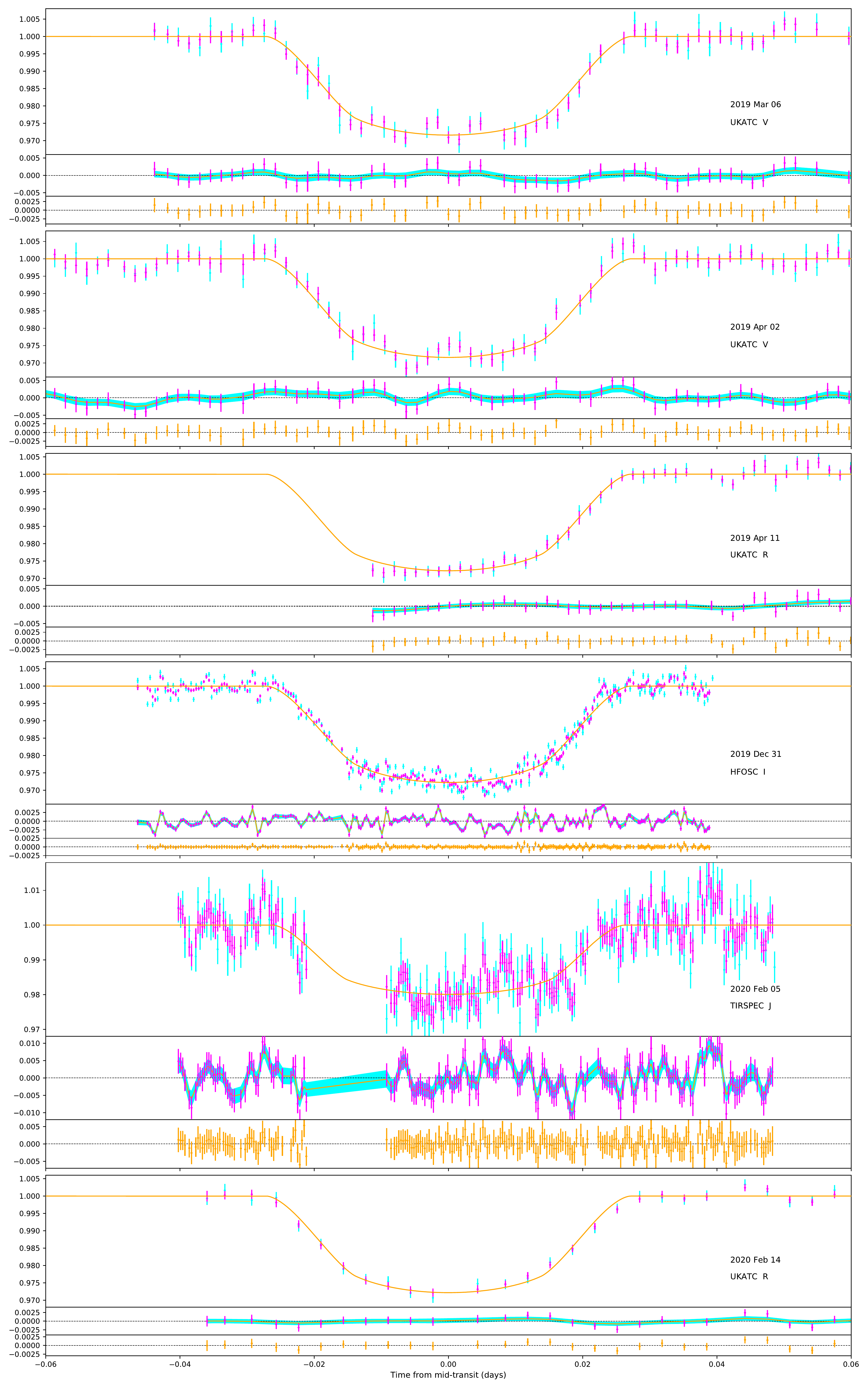}
	\caption{Same as Figure \ref{fig:fig1}, but for WASP-43 b}
	\label{fig:fig3}
\end{figure*}

\begin{figure*}[t]
	\centering
	\includegraphics[width=0.8\linewidth]{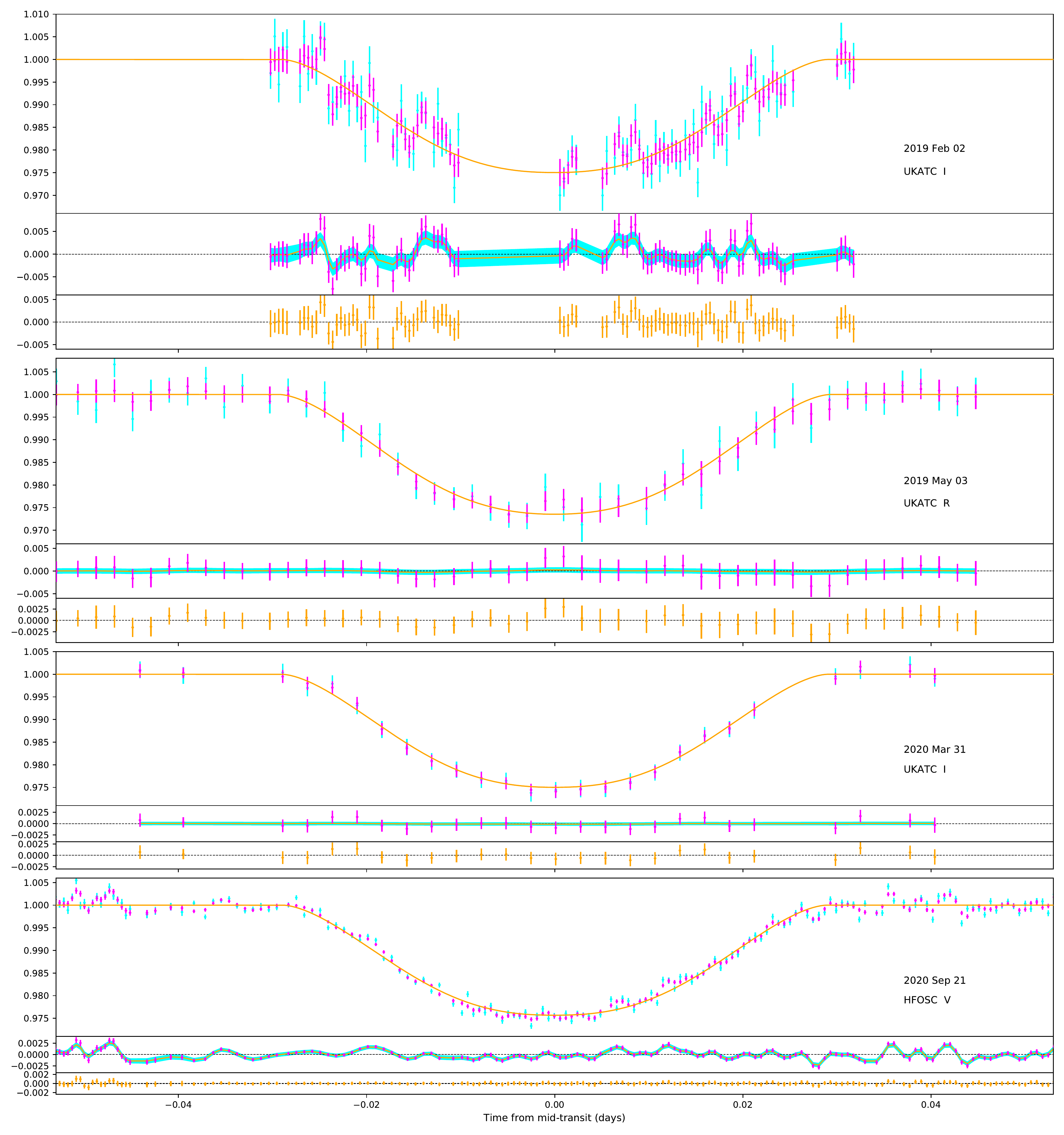}
	\caption{Same as Figure \ref{fig:fig1}, but for TrES-3 b}
	\label{fig:fig4}
\end{figure*}

\begin{figure*}[t]
	\centering
	\includegraphics[width=0.8\linewidth]{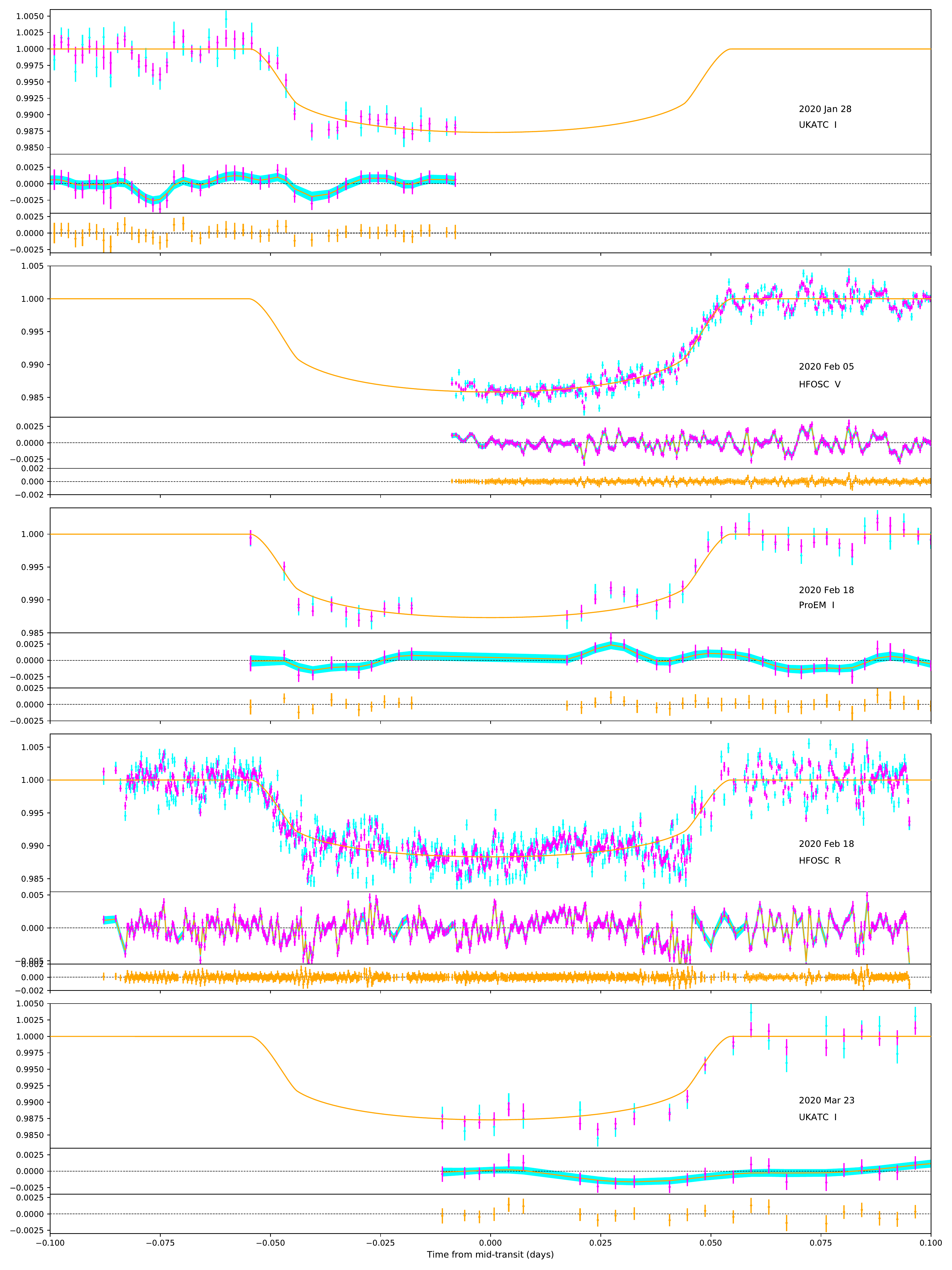}
	\caption{Same as Figure \ref{fig:fig1}, but for XO-2 N b}
	\label{fig:fig5}
\end{figure*}

\pagebreak

\section{Target selection and observations}\label{sec:obs}

In this paper we report the follow-up of five hot-Jupiters, e.g., HAT-P-30 b, HAT-P-54 b, WASP-43 b, TrES-3 b and XO-2 N b. The adopted physical properties of these exoplanets and their host-stars are listed in Table \ref{tab:tab1}.

Our photometric observations are conducted using the 2m Himalayan Chandra Telescope (HCT) at Indian Astronomical Observatory, Hanle and the 1.3m J. C. Bhattacharya Telescope (JCBT) at Vainu Bappu Observatory, Kavalur. At HCT, the Hanle Faint Object Spectrograph Camera (HFOSC) has been used for photometric observations in V, R and I bands (Bessel), whereas the TIFR Near Infrared Spectrometer and Imager (TIRSPEC) has been used for the observations in J band. At JCBT, the UKATC optical CCD and the ProEM imagers have been used for the photometric observations at V, R and I bands. We have also made simultaneous observations of a few similar magnitude stars present within the photometric field of view of the target host-stars. These are  used as reference stars for the differential photometry as described in the next section. The details of our observations have been listed in Table \ref{tab:tab2}. Our observations have been optimised for mid-high cadence and high S/N. As can be seen from the table, the observations from JCBT have mean S/N$>$250. On the other hand, the observations from HCT/HFOSC have very-high cadence and mean S/N$>$1000.

\pagebreak

\section{Data reduction and analysis}\label{sec:analysis}

We have reduced the raw photometric data to obtain the transit light curves, which are then processed through several techniques to reduce the noises from various sources. We have then modelled the processed light curves in order to obtain the transit parameters. These parameters are then used to derive a few other physical properties of the exoplanets. For the whole process of data reduction and analysis, we have used our software package, which is written in Python programming language and uses several open-source software libraries. A detailed explanation of our methodology is given in the following subsections.

\pagebreak

\subsection{Data reduction and differential photometry}

Our observational data obtained at each night, consists of a large number of photometric frames which are reduced through our automatic pipeline that uses the standard PyRAF (IRAF) libraries in the back-end. The raw photometric data are first calibrated using the bias and flat frames obtained during the respective nights of observations. Since, all the instruments used in our observations are cooled up to -70\textsuperscript{o}C, the dark noise is found to be negligible. The calibrated frames are used to obtain the flux from our target host-stars and the reference field stars using aperture photometry. We have calculated the photometric noise precisely using the formula:

\begin{equation}
N = \sqrt{f / g + a \times s^2 + a^2 \times s^2 / k}
\end{equation}

where, $f$ is sky-subtracted flux of the target object, $s$ is the standard deviation of the counts on the region of the sky surrounding the object, $g$ is gain of the instrument, $a$ is the area of aperture of the object chosen in square pixels, and $k$ is number of sky pixels. We have converted the time scale to BJD-TDB using the utc2bjd online applet \citep{2010PASP..122..935E}.

The observations from the ground-based telescopes are heavily affected by the varying atmospheric transparency and air-mass effect which can be reduced using the differential photometry method. We have used the flux of the reference field stars with the best S/N and minimum differential fluctuations for the differential photometry of our target host-stars and obtained the photometric transit light-curves.

\subsection{Baseline correction}

The red noise of large temporal scale (i.e., more than the characteristic scale of transit durations), which are mainly due to various instrumental effects and long-term stellar variability, results in a non-flat baseline for the photometric signals. One way to address this noise component is to model it using simple polynomial fits alongside the transits and subtract from the light-curves. However, it risks a potential bias in the estimated parameters due to the large scale nature of these noise components, and an increment in the parameter load in the already heavily populated model parameter space for MCMC sampling (see \ref{subsection:sub2}).

We have, therefore, performed the baseline correction before modelling for the transit signals. We have used linear and quadratic polynomials of time to model the out-of-transit part of the light-curves and chosen the one with the least BIC (Bayesian Information Criterion) \citep{1978AnSta...6..461S} for the baseline correction. Use of only the out-of-transit portion of the light curves for baseline modelling removes the risk of the low-mid temporal scale variations in transit signal influencing the determination of baseline coefficients. Hence, it removes the risk of potential manipulation in the estimated transit parameters. The normalised baseline corrected transit light-curves from our observations are shown in Figure \ref{fig:fig1}-\ref{fig:fig5}.

\subsection{Wavelet Denoising}

The time-uncorrelated noise (white noise) in the photometric signal also consists of the fluctuations in the light curve due to the small spatial scale variability in the transparency of the Earth's atmosphere. The presence of such fluctuations are more evident in the high-cadence photometric observations due to a better temporal resolution. They can severely affect both accuracy and precision of the estimated transit properties from those light curves. The white noise component cannot be totally removed from a signal and should be only reduced cautiously without distorting the informative part of the signal. Hence, instead of smoothing the light curves with some low-pass filter, we used a more robust technique, namely the Wavelet Denoising \citep{Donoho1994IdealDI, 806084, WaveletDenoise2012}. Although wavelet based denoising techniques have been widely  used in image processing and remote sensing in various fields of science and engineering, it is a recent addition in the context of transit photometry and other light curve analysis \citep[e.g.,][]{2018AA...619A..86D, cubillos17, waldmann14}. CS19 applied this technique on the light curves obtained from their observational data and demonstrated that this technique produces no distortion in the transit light curves, but yields better MCMC posterior distributions for the fitted transit parameters. 

Wavelet Denoising consists of mainly three steps: deconstruction of the original signal into wavelet coefficients using discrete wavelet transform, thresholding, and reconstruction of the signal from the thresholded coefficients. We have used the PyWavelets \citep{Lee2019} python package to perform the single level discrete wavelet transform of our photometric light-curves. In this process, we have used the Symlet family of wavelets, which are the least asymmetric modified version of Daubechies wavelets. A single level transform removes the risk of excess denoising. We have calculated the threshold value using the universal thresholding law \citep{Donoho1994IdealDI} given as;

\begin{equation}
Th = \sigma \sqrt{2log_e(N)}
\end{equation}

where $\sigma = |median(D_x)| / 0.6745$, and performed the hard thresholding, where the wavelet coefficients with absolute values less than the threshold value are replaced with it. The advantage behind universal threshold is that the risk of thresholding is small enough to satisfy the requirement of most applications. The threshold coefficients are then used to construct the denoised signal. The transit light curves after Wavelet Denoising are shown in Figure \ref{fig:fig1}-\ref{fig:fig5}.

\begin{deluxetable*}{LCCC}
	\tablecaption{Estimated values of physical parameters for HAT-P-30 b \label{tab:tab3}}
	\tablewidth{0pt}
	\tablehead{\colhead{Parameter} & \colhead{Filter} & \colhead{This work} & \colhead{\citet{2011ApJ...735...24J}}}
	\startdata
	\text{Transit parameters}\\
	b & & 0.8546_{-0.0055}^{+0.0041} & 0.854_{-0.010}^{+0.008}\\
	R_\star/a & & 0.1430_{-0.0012}^{+0.0013} & 0.1348\pm0.0047\\
	R_p/R_\star & V & 0.10753_{-0.00165}^{+0.00180} & 0.1134\pm0.002^{\:\bm\dag}\\
	& R & 0.11346_{-0.00165}^{+0.00193}\\
	& I & 0.11683_{-0.00104}^{+0.00110}\\
	\text{Limb-darkening coefficients}\\
	C_1 & V & 0.329_{-0.020}^{+0.020} & 0.1975^{\:\bm\dag}\\
	& R & 0.329_{-0.020}^{+0.020}\\
	& I & 0.329_{-0.020}^{+0.020}\\
	C_2 & V & 0.280_{-0.020}^{+0.020} & 0.3689^{\:\bm\dag}\\
	& R & 0.280_{-0.020}^{+0.020}\\
	& I & 0.279_{-0.019}^{+0.021}\\
	\text{Derived parameters}\\
	T_{14}\; [hr] & V & 2.186_{-0.022}^{+0.025} & 2.129\pm0.036^{\:\bm\dag}\\
	& R & 2.213_{-0.021}^{+0.025}\\
	& I & 2.223_{-0.021}^{+0.024}\\
	a/R_\star & & 6.994_{-0.062}^{+0.060} & 7.42\pm0.26\\
	i\; [deg] & & 82.982_{-0.087}^{+0.082} & 83.6\pm0.4\\
	M_p\; [M_J] & & 0.712_{-0.031}^{+0.031} & 0.711\pm0.028\\
	M_p\; [M_\earth] & & 226.2_{-9.7}^{+9.8} & 226.0\pm8.9\\
	T_{eq}\; [K] & & 1686.1_{-24.7}^{+24.8} & 1630\pm42\\
	a\; [AU] & & 0.03948_{-0.00169}^{+0.00170} & 0.0419\pm0.0005\\
	R_p\; [R_J] & V & 1.2995_{-0.0481}^{+0.0482} & 1.340\pm0.065^{\:\bm\dag}\\
	& R & 1.3720_{-0.0495}^{+0.0505}\\
	& I & 1.4122_{-0.0480}^{+0.0482}\\
	R_p\; [R_\earth] & V & 14.566_{-0.539}^{+0.540} & 15.02\pm0.73^{\:\bm\dag}\\
	& R & 15.379_{-0.555}^{+0.566}\\
	& I & 15.829_{-0.538}^{+0.541}
	\enddata
	\tablecomments{$^{\:\bm\dag}$ value doesn't corresponds to the mentioned wavelength filter.}
\end{deluxetable*}

\begin{deluxetable*}{LCCC}
	\tablecaption{Estimated values of physical parameters for HAT-P-54 b \label{tab:tab4}}
	\tablewidth{0pt}
	\tablehead{\colhead{Parameter} & \colhead{Filter} & \colhead{This work} & \colhead{\citet{2015AJ....149..149B}}}
	\startdata
	\text{Transit parameters}\\
	b & & 0.7599_{-0.0054}^{+0.0060} & 0.741_{-0.011}^{+0.010}\\
	R_\star/a & & 0.07175_{-0.00072}^{+0.00098} & 0.0697\pm0.001\\
	R_p/R_\star & V & 0.15616_{-0.00214}^{+0.00204} & 0.1572\pm0.002^{\:\bm\dag}\\
	& R & 0.16898_{-0.00082}^{+0.00115}\\
	& I & 0.17516_{-0.00165}^{+0.00176}\\
	\text{Limb-darkening coefficients}\\
	C_1 & V & 0.627_{-0.026}^{+0.043} & 0.4324^{\:\bm\dag}\\
	& R & 0.685_{-0.035}^{+0.017}\\
	& I & 0.630_{-0.028}^{+0.041}\\
	C_2 & V & 0.080_{-0.007}^{+0.007} & 0.2457^{\:\bm\dag}\\
	& R & 0.080_{-0.007}^{+0.006}\\
	& I & 0.079_{-0.006}^{+0.007}\\
	\text{Derived parameters}\\
	T_{14}\; [hr] & V & 1.819_{-0.017}^{+0.021} & 1.797\pm0.017^{\:\bm\dag}\\
	& R & 1.855_{-0.017}^{+0.020}\\
	& I & 1.872_{-0.018}^{+0.021}\\
	a/R_\star & & 13.94_{-0.19}^{+0.14} & 14.34\pm0.22\\
	i\; [deg] & & 86.873_{-0.057}^{+0.046} & 87.040\pm0.084\\
	M_p\; [M_J] & & 0.761_{-0.032}^{+0.032} & 0.760\pm0.032\\
	M_p\; [M_\earth] & & 241.8_{-10.1}^{+10.3} & 242\pm10\\
	T_{eq}\; [K] & & 832.0_{-10.8}^{+10.9} & 818\pm12\\
	a\; [AU] & & 0.03994_{-0.00098}^{+0.00098} & 0.04117\pm0.00043\\
	R_p\; [R_J] & V & 0.9796_{-0.0330}^{+0.0329} & 0.944\pm0.028^{\:\bm\dag}\\
	& R & 1.0611_{-0.0332}^{+0.0333}\\
	& I & 1.0997_{-0.0357}^{+0.0362}\\
	R_p\; [R_\earth] & V & 10.980_{-0.370}^{+0.369} & 10.6\pm0.3^{\:\bm\dag}\\
	& R & 11.894_{-0.372}^{+0.374}\\
	& I & 12.327_{-0.401}^{+0.406}
	\enddata
	\tablecomments{$^{\:\bm\dag}$ value doesn't corresponds to the mentioned wavelength filter.}
\end{deluxetable*}

\begin{deluxetable*}{LCCC}
	\tablecaption{Estimated values of physical parameters for WASP-43 b \label{tab:tab5}}
	\tablewidth{0pt}
	\tablehead{\colhead{Parameter} & \colhead{Filter} & \colhead{This work} & \colhead{\citet{2017AA...601A..53E}}}
	\startdata
	\text{Transit parameters}\\
	b & & 0.6810_{-0.0058}^{+0.0054} & 0.689\pm0.013\\
	R_\star/a & & 0.2182_{-0.0017}^{+0.0018} & 0.2012\pm0.0057\\
	R_p/R_\star & V & 0.16442_{-0.00156}^{+0.00152} & 0.1588\pm0.004^{\:\bm\dag}\\
	& R & 0.16268_{-0.00142}^{+0.00178}\\
	& I & 0.16214_{-0.00121}^{+0.00123}\\
	& J & 0.1375_{-0.0032}^{+0.0035}\\
	\text{Limb-darkening coefficients}\\
	C_1 & V & 0.627_{-0.026}^{+0.041} & 0.66^{\:\bm\dag}\\
	& R & 0.632_{-0.029}^{+0.041}\\
	& I & 0.680_{-0.036}^{+0.021}\\
	& J & 0.643_{-0.037}^{+0.042}\\
	C_2 & V & 0.079_{-0.007}^{+0.007}\\
	& R & 0.079_{-0.006}^{+0.007}\\
	& I & 0.081_{-0.007}^{+0.006}\\
	& J & 0.080_{-0.006}^{+0.007}\\
	\text{Derived parameters}\\
	T_{14}\; [hr] & V & 1.3046_{-0.0086}^{+0.0091} & 1.164\pm0.24^{\:\bm\dag}\\
	& R & 1.3016_{-0.0089}^{+0.0099}\\
	& I & 1.3010_{-0.0094}^{+0.0091}\\
	& J & 1.2583_{-0.0118}^{+0.0102}\\
	a/R_\star & & 4.584_{-0.037}^{+0.036} & 4.97\pm0.14\\
	i\; [deg] & & 81.46_{-0.13}^{+0.12} & 82.109\pm0.088\\
	M_p\; [M_J] & & 1.994_{-0.073}^{+0.072} & 1.998\pm0.079\\
	M_p\; [M_\earth] & & 633.9_{-23.1}^{+22.7} & 635.0\pm25.1\\
	T_{eq}\; [K] & & 1486.3_{-33.5}^{+33.5} & 1426.7\pm8.5\\
	a\; [AU] & & 0.01387_{-0.00016}^{+0.00016} & 0.01504\pm0.00029\\
	R_p\; [R_J] & V & 1.1006_{-0.0598}^{+0.0599} & 1.006\pm0.017^{\:\bm\dag}\\
	& R & 1.0900_{-0.0593}^{+0.0595}\\
	& I & 1.0857_{-0.0586}^{+0.0587}\\
	& J & 0.9199_{-0.0539}^{+0.0546}\\
	R_p\; [R_\earth] & V & 12.336_{-0.670}^{+0.671} & 11.61\pm0.21^{\:\bm\dag}\\
	& R & 12.218_{-0.664}^{+0.667}\\
	& I & 12.169_{-0.657}^{+0.658}\\
	& J & 10.311_{-0.604}^{+0.612}
	\enddata
	\tablecomments{$^{\:\bm\dag}$ value doesn't corresponds to the mentioned wavelength filter.}
\end{deluxetable*}

\begin{deluxetable*}{LCCC}
	\tablecaption{Estimated values of physical parameters for TrES-3 b \label{tab:tab6}}
	\tablewidth{0pt}
	\tablehead{\colhead{Parameter} & \colhead{Filter} & \colhead{This work} & \colhead{\citet{2009ApJ...691.1145S}}}
	\startdata
	\text{Transit parameters}\\
	b & & 0.8317_{-0.0047}^{+0.0040} & 0.84\pm0.01\\
	R_\star/a & & 0.1681_{-0.0020}^{+0.0021} & 0.1687\pm0.0016\\
	R_p/R_\star & V & 0.16483_{-0.00081}^{+0.00076} & 0.1655\pm0.002^{\:\bm\dag}\\
	& R & 0.17226_{-0.00215}^{+0.00192}\\
	& I & 0.16704_{-0.00177}^{+0.00261}\\
	\text{Limb-darkening coefficients}\\
	C_1 & V & 0.454_{-0.027}^{+0.024} & 0.4378\\
	& R & 0.447_{-0.025}^{+0.028}\\
	& I & 0.448_{-0.026}^{+0.027}\\
	C_2 & V & 0.251_{-0.013}^{+0.013} & 0.2933\\
	& R & 0.249_{-0.013}^{+0.014}\\
	& I & 0.250_{-0.013}^{+0.013}\\
	\text{Derived parameters}\\
	T_{14}\; [hr] & V & 1.387_{-0.015}^{+0.015}\\
	& R & 1.405_{-0.016}^{+0.016}\\
	& I & 1.393_{-0.015}^{+0.014}\\
	a/R_\star & & 5.948_{-0.073}^{+0.072} & 5.926\pm0.056\\
	i\; [deg] & & 81.96_{-0.13}^{+0.13} & 81.85\pm0.16\\
	M_p\; [M_J] & & 1.954_{-0.084}^{+0.085} & 1.91_{-0.080}^{+0.075}\\
	M_p\; [M_\earth] & & 621.1_{-26.6}^{+26.9} & 607.030_{-25.425}^{+23.836}\\
	T_{eq}\; [K] & & 1638.3_{-24.0}^{+24.1}\\
	a\; [AU] & & 0.02293_{-0.00067}^{+0.00068} & 0.02282_{-0.00040}^{+0.00023}\\
	R_p\; [R_J] & V & 1.4881_{-0.0768}^{+0.0770} & 1.336_{-0.037}^{+0.031\:\bm\dag}\\
	& R & 1.5555_{-0.0823}^{+0.0830}\\
	& I & 1.5108_{-0.0796}^{+0.0802}\\
	R_p\; [R_\earth] & V & 16.680_{-0.860}^{+0.863} & 14.975_{-0.415}^{+0.347\:\bm\dag}\\
	& R & 17.434_{-0.922}^{+0.930}\\
	& I & 16.935_{-0.892}^{+0.899}
	\enddata
	\tablecomments{$^{\:\bm\dag}$ value doesn't corresponds to the mentioned wavelength filter.}
\end{deluxetable*}

\begin{deluxetable*}{LCCC}
	\tablecaption{Estimated values of physical parameters for XO-2 N b \label{tab:tab7}}
	\tablewidth{0pt}
	\tablehead{\colhead{Parameter} & \colhead{Filter} & \colhead{This work} & \colhead{\citet{2015AA...575A.111D}}}
	\startdata
	\text{Transit parameters}\\
	b & & 0.1993_{-0.0295}^{+0.0239} & 0.287_{-0.055}^{+0.043}\\
	R_\star/a & & 0.1204_{-0.0008}^{+0.0010} & 0.1261\pm0.0017\\
	R_p/R_\star & V & 0.10702_{-0.00062}^{+0.00064} & 0.1049_{-0.00063}^{+0.00059\:\bm\dag}\\
	& R & 0.09797_{-0.00039}^{+0.00052}\\
	& I & 0.10164_{-0.00148}^{+0.00153}\\
	\text{Limb-darkening coefficients}\\
	C_1 & V & 0.503_{-0.032}^{+0.030} & 0.474_{-0.028}^{+0.030\:\bm\dag}\\
	& R & 0.466_{-0.011}^{+0.023}\\
	& I & 0.487_{-0.026}^{+0.036}\\
	C_2 & V & 0.201_{-0.014}^{+0.013} & 0.171_{-0.070}^{+0.067\:\bm\dag}\\
	& R & 0.194_{-0.010}^{+0.015}\\
	& I & 0.199_{-0.013}^{+0.014}\\
	\text{Derived parameters}\\
	T_{14}\; [hr] & V & 2.628_{-0.016}^{+0.018} & 2.7024\pm0.006^{\:\bm\dag}\\
	& R & 2.606_{-0.016}^{+0.018}\\
	& I & 2.615_{-0.017}^{+0.018}\\
	a/R_\star & & 8.308_{-0.072}^{+0.059} & 7.928_{-0.093}^{+0.099}\\
	i\; [deg] & & 88.63_{-0.17}^{+0.21} & 87.96_{-0.34}^{+0.42}\\
	M_p\; [M_J] & & 0.595_{-0.021}^{+0.021} & 0.597\pm0.021\\
	M_p\; [M_\earth] & & 189.1_{-6.8}^{+6.7} & 189.737\pm6.674\\
	T_{eq}\; [K] & & 1308.4_{-14.8}^{+14.9}\\
	a\; [AU] & & 0.0385_{-0.0013}^{+0.0013} & 0.03673\pm0.00064\\
	R_p\; [R_J] & V & 0.9996_{-0.0521}^{+0.0522} & 1.019\pm0.031^{\:\bm\dag}\\
	& R & 0.9155_{-0.0476}^{+0.0477}\\
	& I & 0.9492_{-0.0509}^{+0.0513}\\
	R_p\; [R_\earth] & V & 11.264_{-0.584}^{+0.585} & 11.422\pm0.347^{\:\bm\dag}\\
	& R & 10.262_{-0.534}^{+0.534}\\
	& I & 10.640_{-0.571}^{+0.575}
	\enddata
	\tablecomments{$^{\:\bm\dag}$ value doesn't corresponds to the mentioned wavelength filter.}
\end{deluxetable*}

\begin{deluxetable*}{lccCCC}
	\tablecaption{Mid-transit times and GP regression parameters for each observed transit event \label{tab:tab8}}
	\tablewidth{0pt}
	\tablehead{\colhead{Target Name} & \colhead{Date} & \colhead{Instrument/Filter} & \colhead{Mid-transit time [BJD-TDB]} & \colhead{$\alpha$} & \colhead{$\tau$}}
	\startdata
	HAT-P-30 b & 2020-02-05 & UKATC/V & 2458885.401561_{-0.001481}^{+0.001551} & 0.00054_{-0.00020}^{+0.00019} & 0.0582_{-0.0264}^{+0.0219}\\
	& 2020-02-05 & HFOSC/I & 2458885.401169_{-0.000772}^{+0.000696} & 0.00158_{-0.00009}^{+0.00009} & 0.0263_{-0.0008}^{+0.0010}\\
	& 2020-02-22 & HFOSC/R & 2458902.265187_{-0.000976}^{+0.000822} & 0.00159_{-0.00011}^{+0.00015} & 0.0240_{-0.0011}^{+0.0009}\\
	& 2020-03-10 & UKATC/I & 2458919.128141_{-0.001620}^{+0.002098} & 0.00358_{-0.00042}^{+0.00045} & 0.0650_{-0.0042}^{+0.0055}\\
	& 2020-03-24 & UKATC/V & 2458933.175811_{-0.001794}^{+0.001766} & 0.00036_{-0.00021}^{+0.00024} & 0.0825_{-0.0592}^{+0.0447}\\
	HAT-P-54 b & 2020-01-15 & UKATC/I & 2458864.201261_{-0.000410}^{+0.000374} & 0.00035_{-0.00021}^{+0.00026} & 0.1007_{-0.0437}^{+0.0492}\\
	& 2020-02-03 & UKATC/I & 2458883.204510_{-0.000402}^{+0.000427} & 0.00095_{-0.00039}^{+0.00039} & 0.0642_{-0.0149}^{+0.0212}\\
	& 2020-02-03 & HFOSC/V & 2458883.204556_{-0.000437}^{+0.000331} & 0.00151_{-0.00016}^{+0.00017} & 0.0711_{-0.0050}^{+0.0069}\\
	& 2020-02-22 & HFOSC/R & 2458902.203596_{-0.000308}^{+0.000231} & 0.00091_{-0.00009}^{+0.00011} & 0.0564_{-0.0048}^{+0.0044}\\
	WASP-43 b & 2019-03-06 & UKATC/V & 2458549.296039_{-0.000409}^{+0.000334} & 0.00112_{-0.00023}^{+0.00032} & 0.0581_{-0.0129}^{+0.0178}\\
	& 2019-04-02 & UKATC/V & 2458576.140878_{-0.000322}^{+0.000294} & 0.00150_{-0.00029}^{+0.00022} & 0.0613_{-0.0088}^{+0.0125}\\
	& 2019-04-11 & UKATC/R & 2458585.087944_{-0.000520}^{+0.000496} & 0.00096_{-0.00029}^{+0.00043} & 0.0848_{-0.0242}^{+0.0251}\\
	& 2019-12-31 & HFOSC/I & 2458849.468659_{-0.000166}^{+0.000188} & 0.00171_{-0.00010}^{+0.00010} & 0.0235_{-0.0010}^{+0.0009}\\
	& 2020-02-05 & TIRSPEC/J & 2458885.259449_{-0.000911}^{+0.000789} & 0.00407_{-0.00027}^{+0.00029} & 0.0276_{-0.0017}^{+0.0018}\\
	& 2020-02-14 & UKATC/R & 2458894.207388_{-0.000365}^{+0.000255} & 0.00675_{-0.00028}^{+0.00032} & 0.0744_{-0.0022}^{+0.0029}\\
	TrES-3 b & 2019-02-18 & UKATC/I & 2458533.473769_{-0.000400}^{+0.000382} & 0.00223_{-0.00021}^{+0.00025} & 0.0281_{-0.0032}^{+0.0035}\\
	& 2019-05-03 & UKATC/R & 2458607.351900_{-0.000402}^{+0.000405} & 0.00051_{-0.00025}^{+0.00022} & 0.0582_{-0.0242}^{+0.0248}\\
	& 2020-03-31 & UKATC/I & 2458940.426286_{-0.000281}^{+0.000337} & 0.00030_{-0.00018}^{+0.00021} & 0.0854_{-0.0330}^{+0.0421}\\
	& 2020-09-21 & HFOSC/V & 2459114.150846_{-0.000244}^{+0.000216} & 0.00105_{-0.00009}^{+0.00010} & 0.0327_{-0.0011}^{+0.0010}\\
	XO-2 N b & 2020-01-28 & UKATC/I & 2458877.220952_{-0.000626}^{+0.000550} & 0.00151_{-0.00021}^{+0.00031} & 0.0653_{-0.0084}^{+0.0088}\\
	& 2020-02-05 & HFOSC/V & 2458885.071089_{-0.000316}^{+0.000290} & 0.00101_{-0.00006}^{+0.00006} & 0.0288_{-0.0008}^{+0.0008}\\
	& 2020-02-18 & ProEM/I & 2458898.149337_{-0.000425}^{+0.000431} & 0.00126_{-0.00030}^{+0.00029} & 0.0896_{-0.0120}^{+0.0129}\\
	& 2020-02-18 & HFOSC/R & 2458898.149390_{-0.000237}^{+0.000249} & 0.00180_{-0.00004}^{+0.00005} & 0.0238_{-0.0005}^{+0.0005}\\
	& 2020-03-23 & UKATC/I & 2458932.154930_{-0.000723}^{+0.000927} & 0.00116_{-0.00030}^{+0.00031} & 0.1501_{-0.0413}^{+0.0557}
	\enddata
	\tablecomments{The mid-transit times for the same transit event observed by two different telescopes in three different instances (2020-02-05 for HAT-P-30 b, 2020-02-03 for HAT-P-54 b and 2020-02-18 for XO-2 N b) have been calculated independently.}
\end{deluxetable*}

\subsection{Gaussian Process Regression}\label{subsection:sub1}

The small-mid temporal scale red noise, which are correlated in time, form the major source of the remaining reducible noise components in our transit light curves after the Wavelet Denoising. This noise is primarily due to the small-scale activity of the host stars. To reduce this noise component, we have used the Gaussian Process (GP) regression \citep{2006gpml.book.....R, 2015ApJ...810L..23J, 2019MNRAS.489.5764P, 2020AA...634A..75B} technique.

We have modelled the time-correlated noise in the transit light-curves using GP regression alongside modelling for the transit signals, using the Markov Chain Monte Carlo (MCMC) technique (see Section \ref{subsection:sub2}). In order to treat the photometric noise in our observed light-curves while modelling for time-correlated noise, we have followed the regression formalism as given by \citet{2006gpml.book.....R} for noisy observations. In this process, we have used the Matérn class covariance function with parameter of covariance, $\nu = 3/2$, and two free parameters, e.g., the signal standard deviation $\alpha$ and the characteristic length scale $\tau$, which are used as model parameters in the MCMC sampling.

\subsection{Markov Chain Monte Carlo sampling}\label{subsection:sub2}

In the final phase of our analysis of the transit light-curves, we have modelled them to the analytical transit formulation following \citet{2002ApJ...580L.171M}, which also incorporates the limb darkening effect of the host-star using the quadratic limb-darkening law given by,

\begin{equation}
I(\mu) = 1 - C_1(1-\mu) - C_2(1-\mu)^2
\end{equation}

We have used the MCMC sampling technique for this modelling, and simultaneously modelled for the time-correlated noise in the signal using the GP regression method as explained in section \ref{subsection:sub1}. The model parameters in MCMC includes the transit parameters: (i) mid-transit time $T$, (ii) the impact-parameter $b$, (iii) the scaled stellar radius $R_\star/a$, (iv) the ratio of planet to stellar radius $R_p/R_\star$, (v) the out-of-transit flux $f_{o}$, (vi) the limb darkening coefficients $C_1$ and $C_2$; and (vii) the GP regression coefficients $\alpha$ and $\tau$. The orbital periods have been kept constant and are adopted from previous studies as given in Table \ref{tab:tab1}. We have followed \citet{2010AA...510A..21S} for the prior values of the quadratic limb-darkening coefficients.

We have modelled all the transit light-curves corresponding to each of our target exoplanets simultaneously keeping the transit independent parameters the same for all transit events of the same target and keeping the wavelength dependent parameters same across those transit events observed using the same photometric filters. We have used the Metropolis-Hastings algorithm \citep{1970Bimka..57...97H} in MCMC sampling and used the modified marginal likelihood function due to the inclusion of the GP regression technique \citep{2006gpml.book.....R}. In MCMC sampling for each of our targets, we have used 30 walkers and 100,000 iterations for each independent walker. Such a large volume of sampling is done in order to handle the large number of model parameters and high volume of photometric data under simultaneous modelling of multiple transit events. The aim has been to assess the robustness of the convergence of the sampling parameters such that a high accuracy in the derived parameters is achieved. Such a large computation is practically impossible without the incursion of parallel computing. For this purpose, our software package for modelling and data analysis has been facilitated with parallel computing using the Python-based \textit{multiprocessing} library. We have used the high performance computing facility (NOVA) at Indian Institute of Astrophysics (IIA) for our numerical computation. We have discarded the first 10,000 iterations for each walkers as burn-in and accepted the rest 90,000 iterations for obtaining the posterior distributions. We have shown the corner diagrams depicting the posterior distributions of the transit parameters and the GP regression coefficients from our MCMC sampling in the appendix \ref{app}.

We present the best-fit transit models, best-fit GP regression model, and the mean residual fluxes in Figure \ref{fig:fig1}-\ref{fig:fig5}. The estimated values of the physical properties for our target exoplanets with 1$\sigma$ error have been given in Table \ref{tab:tab3}-\ref{tab:tab7} and the estimated values of mid-transit times along with the GP regression coefficients for each transit light-curves have been given in Table \ref{tab:tab8}.

\subsection{Derived parameters}

The posterior distribution of the transit parameters from the MCMC sampling along with the adopted physical properties of the target exoplanets and the host stars as given in Table \ref{tab:tab1} are used to estimate the values of some other physical parameters.

We have estimated the transit duration $T_{14}$ for each of our targets at each photometric band using the relation (CS19)

\begin{equation}
T_{14} = \frac{P}{\pi}\sin^{-1} \begin{pmatrix}
\frac{\sqrt{(1 + R_p / R_\star)^2 - b^2}}{\sqrt{(a / R_\star)^2 - b^2}}.
\end{pmatrix}
\end{equation}

The inclination angle of the planetary orbit $i$ is estimated using the relation

\begin{equation}
i = \cos^{-1} \begin{pmatrix}
\frac{bR\star}{a}
\end{pmatrix}
\end{equation}

We have estimated the mass $M_p$ and the equilibrium temperature $T_{eq}$ of the target exoplanets using the relations (CS19),

\begin{equation}
M_p = M_\star^{2/3} \begin{pmatrix}
\frac{P}{2 \pi G}
\end{pmatrix}^{1/3}
\frac{K_{RV}}{\sin i}
\end{equation}

and

\begin{equation}
T_{eq} = T_{eff} \sqrt{\frac{R_\star}{2a}}
\end{equation}

For all the cases, we have assumed circular orbits, zero Bond albedo, and full re-circulation of the incident stellar flux over the planetary surface. The estimated values of all the derived physical properties for each of our targeted exoplanets are given in the Table \ref{tab:tab3}-\ref{tab:tab7}.

\section{Results and Discussion}\label{sec:dis}

Our study involves the photometric follow up observations of five hot-Jupiters using a 1.3m (JCBT) and a 2m (HCT) class telescopes. Table \ref{tab:tab1} shows that the observations from HCT have a better S/N and time-cadence. This can be attributed to the large aperture of HCT, short readout time of the new CCD installed at the back-end (HFOSC2), and good observational condition at the site of IAO. However, the observations from JCBT also show good S/N and have been extremely useful in complementing the observations from the HCT. Since the number of nights allocated at each telescope is limited, we use both the telescopes for our observations. 

Figure \ref{fig:fig1}-\ref{fig:fig5} demonstrate that the Wavelet Denoising technique is a useful tool for the reduction of the time-uncorrelated noise components and outliers from the transit light curves. Also, we have avoided excessive smoothing of the light-curves by cautiously choosing the decomposition levels and threshold levels while using this denoising technique.

The GP regression method has very effectively modelled the correlated noise components in the transit signal as shown in  Figure \ref{fig:fig1} - \ref{fig:fig5}. The origin of this correlated noise component is not only attributed to the low-scale variability of the exoplanet host-stars, but also to that of the reference stars used for differential photometry. Since, the reference stars are different in most of the cases, due to various observational reasons, the amplitude of the correlated noises are different for the same target host star. Similarly, the correlated noise components are negligible for some of the low-cadence observations from JCBT. Our analysis shows the efficiency of GP regression technique in reducing the correlated noise components irrespective of their origin and amplitude.

Our MCMC sampling includes a large number of sampling points, which is helpful to remove any bias from the choice of prior values in the MCMC, thereby resulting in an more accurate estimation of the free parameters. This  is evident from the near-normal distribution of most of the model-parameter posterior distributions as shown in the figures in appendix \ref{app}. As can be seen from the estimated values of the mid-transit times of our observed transit events, the independently estimated mid-transit times for the same transit event observed from the two different telescopes (HCT and JCBT) in three different instances [2020-02-05 for HAT-P-30 b, 2020-02-03 for HAT-P-54 b and 2020-02-18 for XO-2 N b] are in perfect agreement with each other, with the error-bar in the estimated value from the HCT observation being small due to a better S/N. This proves the robustness of our analysis method in the estimation of these properties. 

Due to the multi-band observations of the transit events, we have been able to estimate the wavelength dependent radii of our target exoplanets with good accuracy at each band. Even though the site conditions at IAO allows observation in the infrared wavelengths, the S/N of the data observed using the TIRSPEC instrument at the back-end of HCT is significantly low, especially in the context of transit photometry. However, we managed to observe one frame in the J-band for WASP-43 b. Apart from this, all of the targets have been observed in V-, R- and I-bands. We will further analyse these results to check if any information can be excavated about the atmospheres of these planets with the help of the models we have developed for transmission spectra \citep{chakrabarty20, sengupta20}.

We have compared the estimated values of the physical parameters for the target exoplanets in our study with those from the previous studies (given in Tables \ref{tab:tab3}-\ref{tab:tab7}) to understand the effectiveness of the critical noise reduction and modelling algorithm used in our study, and the improvement in the accuracy and precision in the parameter estimation. While for most of the parameters, the estimated values are in good accordance with those from previous studies, small differences in a few parameter values can be attributed to the improvement in the quality of the transit signals due to the use the noise reduction algorithms, thereby improving the accuracy in parameter estimation. Also, the precision in the estimated values of the transit parameters and the parameters which are directly derived from them have been improved up to 4 times compared to the previous studies, which is significant in the context of transit photometric studies of exoplanets. This improvement in the uncertainties in estimated parameters is due to a better S/N in our follow-up observations and further reduction of various noise components. On the other hand, the estimated values of other derived parameters, which are dependent upon the stellar properties of the host-stars for their estimation (adopted from the previous studies and given in Table \ref{tab:tab1}), have seen no significant improvements in their precision. This is due to the large uncertainties embedded with those adopted values of stellar parameters.

\section{Conclusion}\label{sec:con}

In this study, we have performed new multi-band transit photometric follow-up of five hot-Jupiters, HAT-P-30 b, HAT-P-54 b, WASP-43 b, TrES-3 b and XO-2 N b, using the 2m Himalayan Chandra Telescope (HCT) at IAO, Hanle and the 1.3m J. C. Bhattacharya Telescope (JCBT) at VBO, Kavalur. Taking the advantage of the larger apertures of these telescopes compared to those used for previous studies of these exoplanets, we have obtained transit light-curves with better S/N.

Our critical noise treatment analysis employs the Wavelet Denoising technique for the reduction of time-uncorrelated noise components from the photometric light curves without potential loss of signal due to transit origins and the Gaussian Process regression technique to effectively model and compensate for the time-correlated noise in the photometric signal simultaneously along with the transit modelling. We have used the Markov Chain Monte Carlo (MCMC) sampling technique by adopting the Metropolis-Hastings algorithm. 

Due to the high S/N photometric follow-up observations as well as the adopted highly optimised noise reduction and analysis techniques, the estimated physical parameters from our study have a better accuracy and overall precision. Also, the follow-up using multiple photometric bands have enabled us to estimate accurately, the wavelength dependent physical properties which can be used as an outset for the high-resolution atmospheric characterisation of these exoplanets in future.

Our ongoing project of follow-up studies of the exoplanets can be extended for the study of other existing exoplanets in  future. Also, the critical noise analysis algorithm can be used for the analysis of transit light curves from the existing and upcoming global exoplanet survey missions, such as the TESS (Transiting Exoplanet Survey Satellite), for the estimation of physical properties of the detected exoplanets with better accuracy.

We are thankful to the anonymous reviewer for a critical reading of the manuscript and for providing many useful comments and suggestions. We thank to the supporting staffs at the Indian Astronomical Observatory (IAO), Hanle, the Center For Research and Education in Science and Technology (CREST), Hosakote, and the Vainu Bappu Observatory (VBO), Kavalur. Some of the computational results reported in this work were performed on the high performance computing facility (NOVA) of IIA, Bangalore. We are thankful to the computer division of Indian Institute of Astrophysics for the help and co-operation extended for the present project. We have used PyRAF for most of the tasks of reduction and photometry. PyRAF is a product of the Space Telescope Science Institute, which is operated by AURA for NASA.

\pagebreak

\bibliography{ms}{}
\bibliographystyle{aasjournal}

\appendix
\restartappendixnumbering

\section{Posterior distributions from MCMC sampling}\label{app}

\begin{figure}[h]
	\includegraphics[width=\linewidth]{f6.pdf}
	\caption{Corner diagram depicting the posterior distributions of the transit parameters and the GP regression coefficients from MCMC sampling for HAT-P-30 b.} 
	\label{fig:fig6}
\end{figure}

\begin{figure*}
	\includegraphics[width=\linewidth]{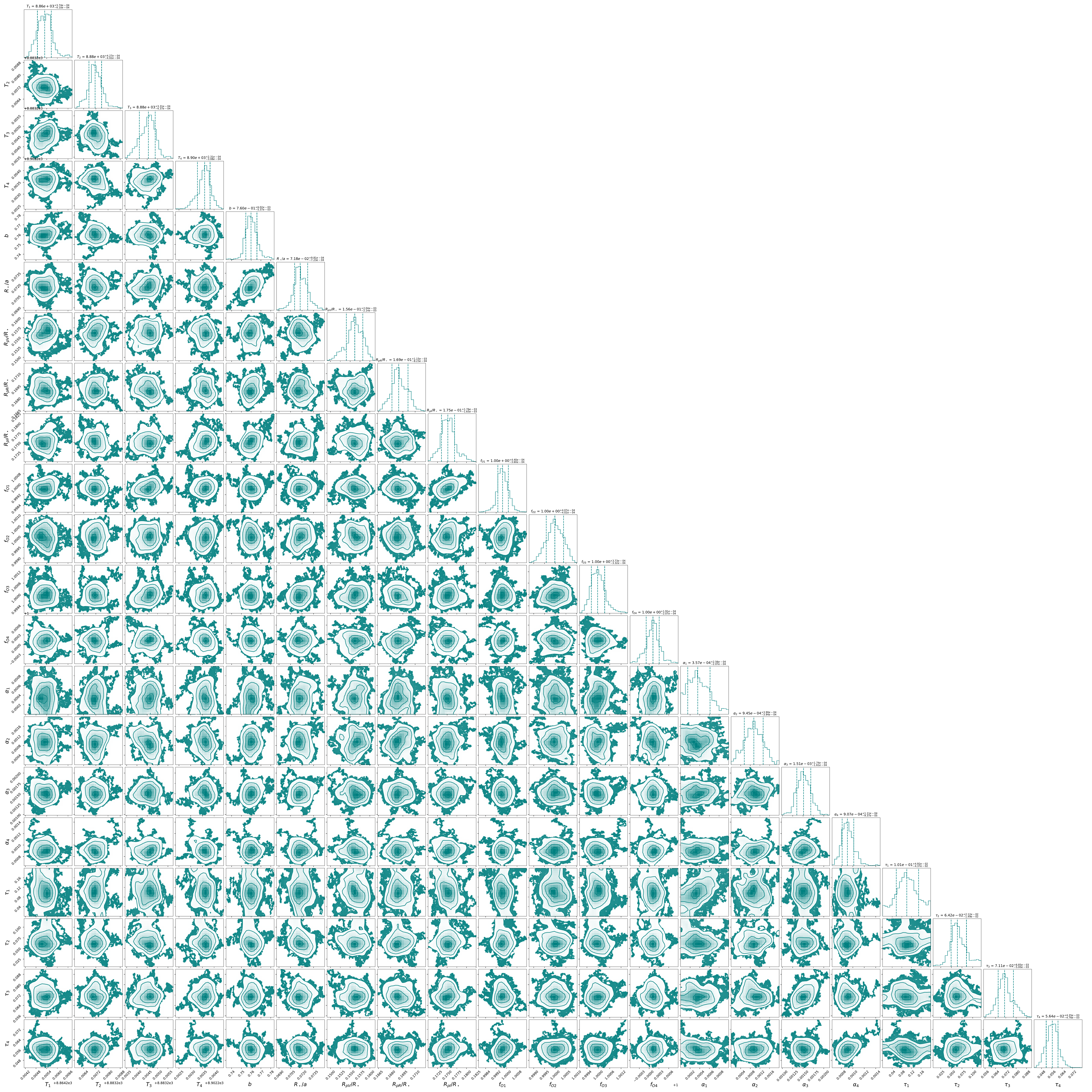}
	\caption{Same as Figure \ref{fig:fig6}, but for HAT-P-54 b}
	\label{fig:fig7}
\end{figure*}

\begin{figure*}
	\includegraphics[width=\linewidth]{f8.pdf}
	\caption{Same as Figure \ref{fig:fig6}, but for WASP-43 b}
	\label{fig:fig8}
\end{figure*}

\begin{figure*}
	\includegraphics[width=\linewidth]{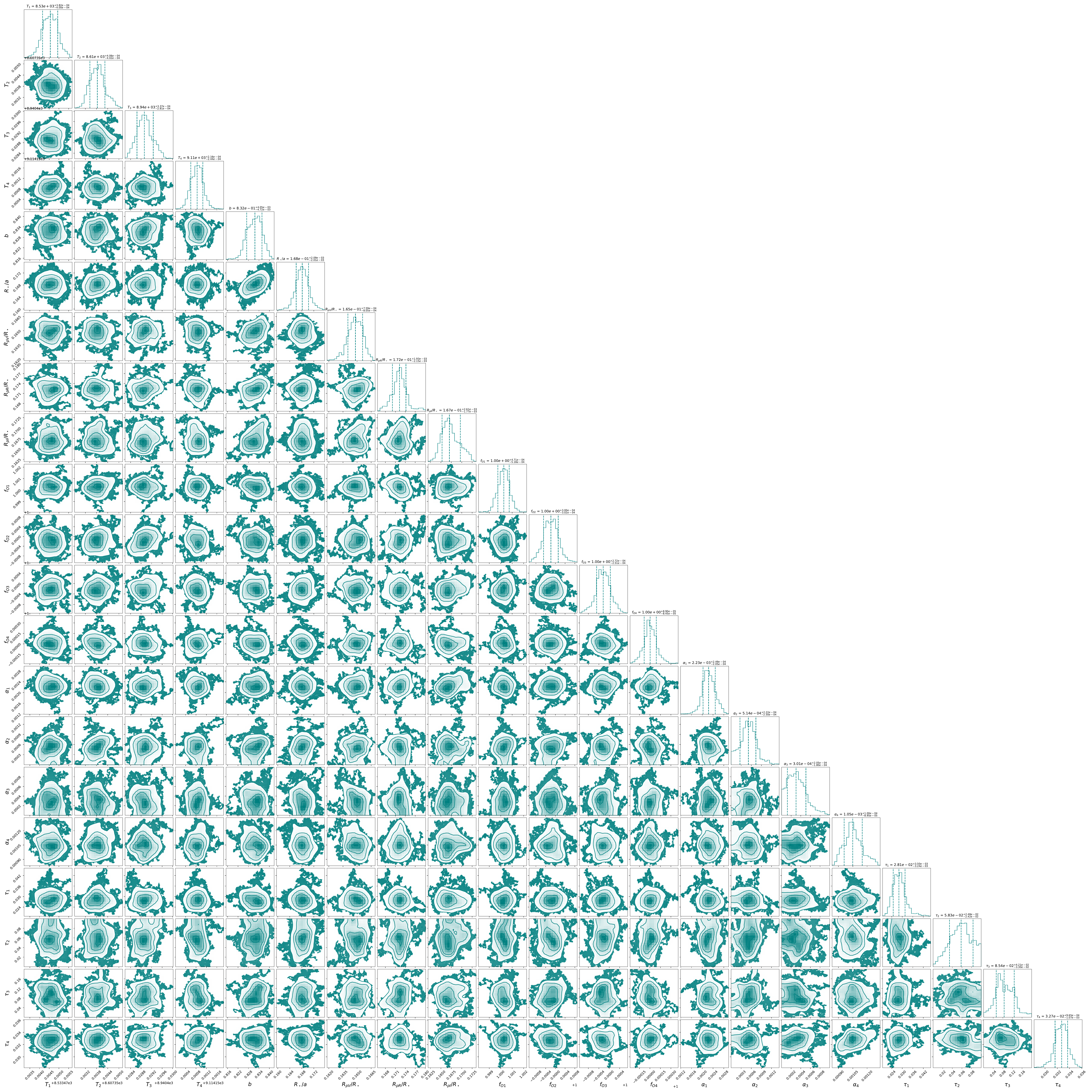}
	\caption{Same as Figure \ref{fig:fig6}, but for TrES-3 b}
	\label{fig:fig9}
\end{figure*}

\begin{figure*}
	\includegraphics[width=\linewidth]{f10.pdf}
	\caption{Same as Figure \ref{fig:fig6}, but for XO-2 N b}
	\label{fig:fig10}
\end{figure*}

\end{document}